\documentclass[%
 reprint,
superscriptaddress,
preprintnumbers,
amsmath,amssymb,
 aps,
prl
]{revtex4-1}

\usepackage{upgreek}
\usepackage[colorlinks=true,linkcolor=blue,citecolor=blue,urlcolor=blue]{hyperref}
\usepackage{graphicx}
\usepackage{mathtools}
\def\Fig#1{Fig.~\ref{#1}}
\def\Eq#1{Eq.~\eqref{#1}}

\usepackage{relsize,exscale,comment}

\usepackage{upgreek}
\usepackage{bm}
\usepackage[caption=false]{subfig}

\begin{document} 
\preprint{CERN-TH-2022-046}

\title{Medium-enhanced \texorpdfstring{$c\bar{c}$}{ccbar} radiation}

\author{Maximilian Attems}
\email{maximilian.attems@cern.ch}
\affiliation{Theoretical Physics Department, CERN, 1211 Geneva 23, Switzerland}
\author{Jasmine Brewer}
\email{jasmine.brewer@cern.ch}
\affiliation{Theoretical Physics Department, CERN, 1211 Geneva 23, Switzerland}
\author{Gian Michele Innocenti}
\email{gian.michele.innocenti@cern.ch}
\affiliation{Experimental Physics Department, CERN, 1211 Geneva 23, Switzerland}
\author{Aleksas Mazeliauskas}
\email{aleksas.mazeliauskas@cern.ch}
\affiliation{Theoretical Physics Department, CERN, 1211 Geneva 23, Switzerland}
\author{Sohyun Park}
\email{sohyun.park@cern.ch}
\affiliation{Theoretical Physics Department, CERN, 1211 Geneva 23, Switzerland}
\author{Wilke van der Schee}
\email{wilke.van.der.schee@cern.ch}
\affiliation{Theoretical Physics Department, CERN, 1211 Geneva 23, Switzerland}
\author{Gregory Soyez}
\email{gregory.soyez@ipht.fr}
\affiliation{
IPhT, Universit\'e Paris-Saclay, CNRS UMR 3681, CEA Saclay, F-91191 Gif-sur-Yvette, France
}
\author{Urs Achim Wiedemann}
 \email{urs.wiedemann@cern.ch}
\affiliation{Theoretical Physics Department, CERN, 1211 Geneva 23, Switzerland}

\begin{abstract} 
We show that the same QCD formalism that accounts for the suppression of high-$p_T$  hadron and jet spectra in heavy-ion collisions
predicts medium-enhanced production of $c\bar{c}$ pairs in jets. We demonstrate that this phenomenon, which cannot be accessed by traditional jet-quenching observables, can be directly observed using $D^0\bar{D}^0$-tagged jets in nuclear collisions.
\end{abstract}

\maketitle

\noindent
{\bf Introduction.}
In ultra-relativistic heavy-ion collisions, heavy flavor quarks (charm and beauty) are produced in high-momentum transfer processes, which are calculable in perturbative QCD (pQCD)~\cite{Cacciari:2012ny}. As these heavy quarks traverse the Quark Gluon Plasma (QGP) created in the collision, they probe its properties. Over the lifetime of the QGP, charm and beauty quarks are stable and can be tagged experimentally. This makes them ideally suited for studying quark propagation in the QCD plasma~\cite{Andronic:2015wma}.

\begin{figure}
\includegraphics[width=\linewidth,trim=0 50 0 40,clip]{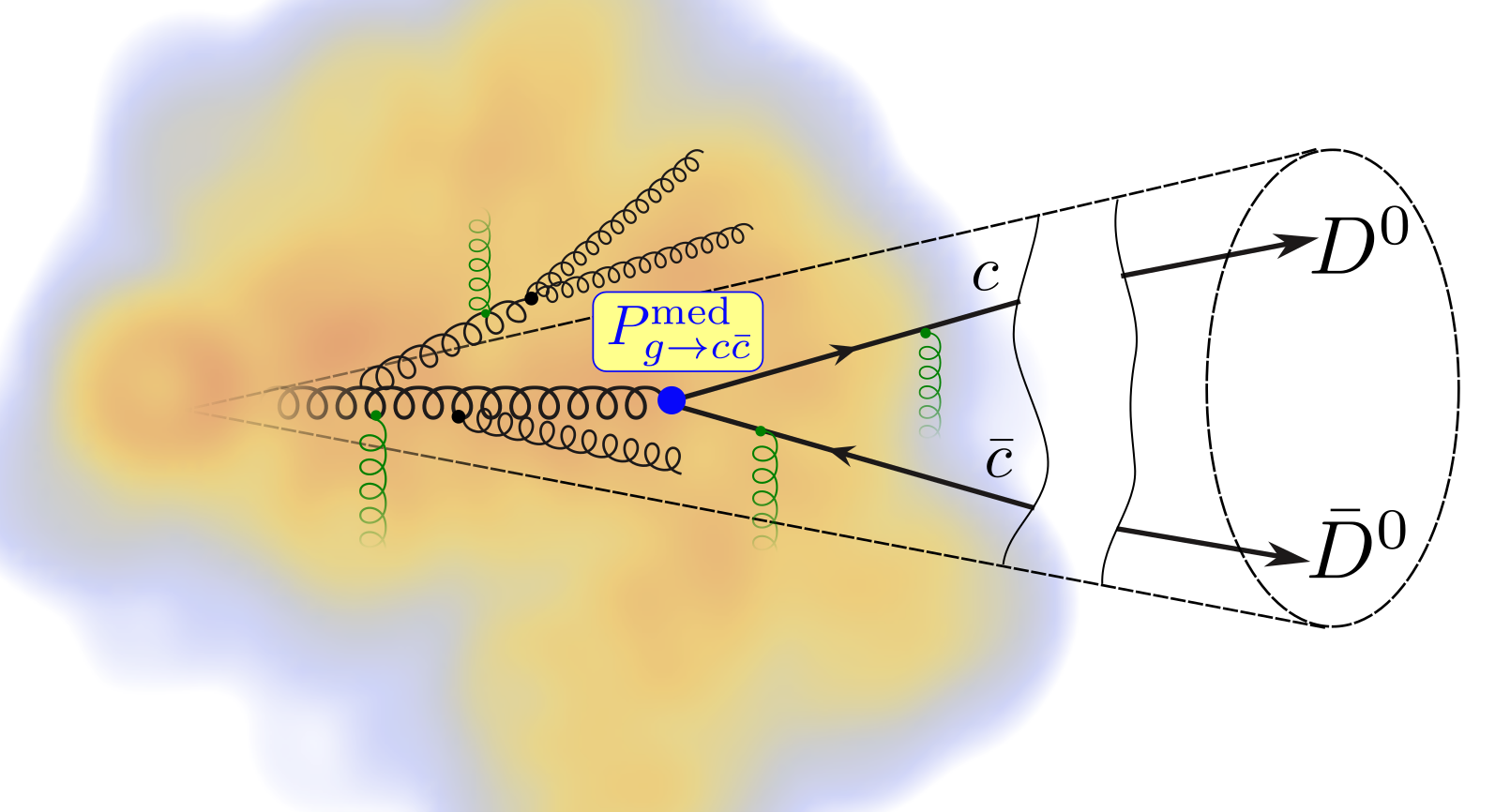}
\caption{Illustration of a parton shower containing a $g \to c\bar{c}$ splitting.
This $c\bar{c}$ radiation is enhanced due to the QGP, which
can be tested by measuring $D^0\bar{D}^0$ pairs inside jets.
}
\label{cartoon}
\end{figure}

Theory predicts that energetic quarks and gluons (partons) lose energy while traversing the QGP.  This \emph{jet quenching} phenomenon is observed in heavy-ion collisions as a generic suppression of high-$p_T$ single inclusive hadron and jet spectra~\cite{Connors:2017ptx}. 
The suppression is understood as a combination of elastic energy loss and radiative energy loss from medium-induced gluon radiation~\cite{Baier:1996kr,Zakharov:1996fv,Dokshitzer:2001zm}.
This theory framework explains the generic suppression of  high-$p_T$ hadron and jet spectra, and also predicts the enhancement of $q\bar{q}$ pairs in the phase space region in which medium-induced radiation occurs. However, in contrast to the suppression patterns that gave ``jet quenching" its name, such characteristic enhancement has not been tested experimentally so far. Here, we show that heavy quarks provide a unique avenue to measure this effect because gluon splittings into charm quarks can be tagged experimentally in jets containing a pair of heavy-flavor mesons, see Fig.~\ref{cartoon}.

In hadronic collisions, charm is produced in $c\bar{c}$-pairs of squared invariant mass  $Q^2= \left( p_\text{c} + p_{\bar{\text{c}}} \right)^\mu \left( p_\text{c} + p_{\bar{\text{c}}} \right)_\mu$, 
bounded by the charm quark mass $m_c$ and the partonic center of mass energy $\hat{s}$, $4 m_c^2 \leq Q^2 \leq \hat{s}$.
Pairs can be produced at high $Q^2$ (typically through hard scattering or initial-state radiation) or at low $Q^2$ (often from gluon splitting), see e.g.~\cite{Cacciari:2012ny,Carrer:2003ac}. Up to now most studies have focused on the gluon radiation $c\to c\, g$ that can reduce the charm momentum but leaves its yield unaffected~\cite{Dokshitzer:2001zm}.  In the present Letter, we emphasize that radiative parton energy loss predicts a qualitatively different phenomenon: For small $Q^2$, the $g \to c\bar{c}$ splitting receives significant medium-modifications that increase the yield of $c\bar{c}$-pairs within high-$p_T$ jets.

In the collinear limit $Q^2 \ll \hat{s}$, partonic cross sections for $c\bar{c}$ production factorize. For instance, the cross section for $gg \rightarrow c\bar{c}X$ can be written as
\begin{equation}
	\hat{\sigma}^{g\, g \to c\, \bar{c}\, X} \,
	\xrightarrow{Q^2 \ll \hat{s}}\,  \hat{\sigma}^{g\, g \to g\, X} \, \frac{\alpha_s}{2\pi}\, \frac{1}{Q^2}\, P_{g \to c\bar{c}}(z) \,.
	\label{eq1}
\end{equation}
The $g\to c\bar{c}$ splitting function depends on the momentum fraction $z$ carried by the charm quark
and the virtuality $Q$ of the gluon,
\begin{equation}
 P^\text{vac}_{g \to c\bar{c}} =\frac{1}{2} \left(z^2 + (1-z)^2 + \frac{2\, m_c^2}{Q^2}\right)\, . 
\label{eq2}
\end{equation}
We use $E_g$ to denote the gluon energy and $2{\bm \upkappa}$ the relative $c\bar{c}$ pair momentum transverse to the direction of the gluon. 
For collinear splittings $\bm{\upkappa} \ll z E_g, (1-z)E_g$ the squared gluon virtuality is
$Q^2 = \tfrac{m_c^2 + \bm{\upkappa}^2}{z(1-z)}$.

For sufficiently high gluon energy, the gluon is boosted with respect to the QGP by a Lorentz factor $\gamma = E_g/Q$, and the formation time $\sim 1/Q$ of the $c\bar{c}$ pair is time-delayed by $\gamma$ and becomes $\sim E_g/Q^2$ in the rest frame of the QGP.
For pairs within a jet we are restricted to small $Q^2$ and the formation time can be comparable to the medium size. For instance, for $E_g = 100\,$GeV and $Q^2=4 m_c^2$, one finds $E_g/Q^2 = 3.1\,$fm$/c$. We hence expect a characteristic enhancement of $c\bar{c}$ pairs within the same jet depending on energy and the virtuality of the pair.

 {\bf Medium-modified $g\to c\bar{c}$ splitting function.}
 For the $q \to q\, g$ and $g \to g\, g$ leading-order splitting functions, medium modifications have been calculated in 
the BDMPS-Z formalism~\cite{Baier:1996kr,Zakharov:1996fv,Wiedemann:2000za}  and in several related set-ups~\cite{Gyulassy:2000er,Wang:2001ifa}.
 These calculations resum the effects of multiple interactions between the QCD plasma and the splitting process in a close-to-eikonal formulation, where the energy of the parent parton is much larger than any transverse momentum or mass scale. For the $g\to c\bar{c}$ splitting function, medium-modifications have been calculated to first order in opacity~\cite{Kang:2016ofv,Attems:2022ubu} as well as in the
 BDMPS-Z path-integral formalism where they read~\cite{Attems:2022ubu}
\begin{align}
&\left(\frac{1}{Q^2}\, P_{g \to c\, \bar{c}} \right)^{\rm tot}
= 	\left(\frac{1}{Q^2}\, P_{g \to c\, \bar{c}} \right)^{\rm vac} +
\left(\frac{1}{Q^2}\, P_{g \to c\, \bar{c}} \right)^{\rm med}
	\nonumber \\
&		= 2\, \mathfrak{Re}\, \frac{1}{8\, E_g^2}\, \int_0^{\infty} dt \int_t^{\infty} d\bar{t} \int d{\bf r}
		\nonumber \\
		 & \times \, 
    e^{i\frac{m_c^2}{2E_g z(1-z)} (t-\bar{t}) - \epsilon |t| - \epsilon |\bar{t}|}\,
     e^{ - \frac{1}{4} \int_{\bar{t}}^\infty d\xi\, \hat{q}(\xi,z)\, {\bf r}^2}\, 
		  e^{-i\, \bm{\upkappa} \cdot{\bf r}}
		  \label{eq3}\\		  		
		&\times \left[ \frac{ m_c^2}{z(1-z)}+  \frac{z^2 + (1-z)^2}{z(1-z)}\frac{\partial}{\partial {\bf x}}\cdot \frac{\partial}{\partial {\bf r}}
		    \right] \, {\cal K}\left[{\bf x}=0,t;{\bf r},\bar{t}\right]\, .
		   \nonumber
\end{align}
This expression is written in the limit of a large number of colors.
It has the space-time interpretation of a gluon that splits into a $c\bar{c}$ pair at longitudinal positions $t$ ($\bar{t}$) in amplitude (complex conjugate amplitude). In the multiple soft scattering approximation,
${\cal K}\big[{\bf x},t;{\bf r},\bar{t}\big] $ 
is the path-integral of a harmonic oscillator with imaginary potential
\begin{equation}
{\cal K}\big[{\bf x},t;{\bf r},\bar{t}\big] = \int_{{\bm{\uprho}}(t)={\bf x}}^{{\bm{\uprho}}(\bar{t})={\bf r}}
{\cal D}{\bm{\uprho}}\, e^{ i \int_t^{\bar t} d\xi\, \left(  \frac{E_g z(1-z)}{2}  \dot{\bm{\uprho}}^2 - \frac{\hat{q}(\xi,z)\,\bm{\uprho}^2}{4\, i} \right)}\, .\nonumber 
\end{equation}

It describes how the $c\bar{c}$ dipole grows from transverse size ${\bf x}=0$ at $t$ to size ${\bf r}$ at $\bar{t}$. Equation \eqref{eq3} depends on the kinematics of the splitting and on a single medium property, the quenching parameter  $\hat{q}(\xi,z)$ along the parton trajectory in the medium.

 \begin{figure}
\includegraphics[width=\linewidth]{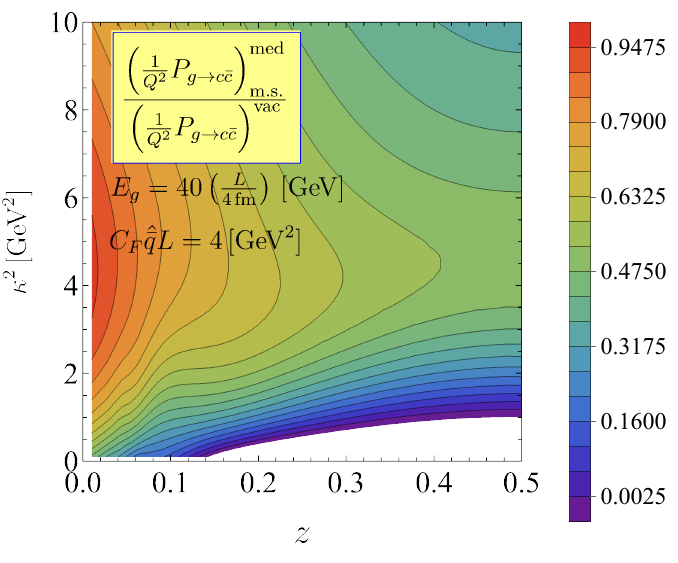}
  \caption{The ratio of 
  medium-modification over vacuum splitting $P^\text{med}_{g\to c\bar{c}}/P^\text{vac}_{g\to c\bar{c}}\equiv P^\text{tot}_{g\to c\bar{c}}/P^\text{vac}_{g\to c\bar{c}}-1 $ shows an enhancement of nearly 100\% over vacuum baseline in a significant phase space region (see \Eq{eq:weight}).
  }
\label{fig2}
\end{figure} 
Both adjoint and fundamental color configurations contribute in a $z$-dependent way to the interaction of the $g\to c\bar{c}$ splitting with the QGP~\cite{Attems:2022ubu}. This dependence can be factorized  from the quenching parameter, $\hat{q}(\xi,z) \equiv (C_F-C_A z(1-z))\, \hat{\bar{q}}$. For small $z$, the quenching parameter is hence $\hat{q}=C_F \hat{\bar{q}}$, in contrast to $\hat{q}_A \equiv C_A\, \hat{\bar{q}}$ for $q\to q\, g$ and  $g\to g\, g$, for which the leading medium modification is from the scattering of the emitted gluon~\cite{Baier:1996kr,Zakharov:1996fv,Wiedemann:2000za}.

To estimate $\hat{\bar{q}}$ 
we considered recent constraints~\cite{JET:2013cls,Andres:2016iys,Andres:2019eus,Huss:2020whe,JETSCAPE:2021ehl} on $\hat{q}_A$ from heavy-ion data to determine the momentum transferred from the medium
\begin{equation}
	\langle q^2\rangle_\text{med} =C_F \int_{\tau_i}^{\tau_f} d\xi\,  \hat{\bar{q}}(\xi) \, .
	\label{eq4}
\end{equation}
Here, $\tau_{i/f}$ denote the initial and final time within which jet-medium interactions occur. Explicit calculation of the line integrals for the jet quenching models in~\cite{Huss:2020whe} yields consistently $4\, \text{GeV}^2 < \langle q^2\rangle_\text{med} < 8\, \text{GeV}^2$ for the 0--5\% most central PbPb collisions at $\sqrt{s_{NN}}=5.02\, \text{TeV}$. 

Figure~\ref{fig2} shows numerical results of the medium-modification of \Eq{eq3} for a momentum transfer 
$\hat{q}L = 4\, \text{GeV}^2$ \footnote{Since \Eq{eq3} depends on only two combinations of 
$\hat{q}$, $L$ and $E_g$, only two dependencies need to be specified in this figure.}.
Medium-induced transverse momentum broadening leads to a characteristic enhancement at the scale 
$\bm{\upkappa}^2 \sim \hat{q}\, L$ while depleting the vacuum distribution of $c\bar{c}$ pairs at very small relative momenta $\bm{\upkappa}^2$. This can be understood in terms of transverse Brownian motion of the charm quarks in the medium, which pushes the collinear pairs to larger $\bm{\upkappa}^2$.
Compared to the vacuum splitting function, the medium-modified splitting can be almost 100\% larger in a range of intermediate $\bm{\upkappa}^2$ for $\hat{q}L = 4\, \text{GeV}^2$. 

{\bf Tracing $g\to c\bar{c}$ via $D^0\bar{D}^0$ pairs in jets.} 
We now discuss one strategy for testing the enhanced $g\to c\bar{c}$ radiation experimentally. To this end, we simulate $pp$ collisions at $\sqrt{s_{NN}} = 5.5$~TeV in \textsc{Pythia} 8.3 (Monash tune)~\cite{Sjostrand:2014zea} with initial state radiation (ISR) off and use FastJet~\cite{Cacciari:2011ma} to reconstruct anti-$k_t$~\cite{Cacciari:2008gp} jets with jet radius $R=0.4$. We select jets with exactly one $D^0\bar{D}^0$ pair, which leads to a high-purity sample of jets in which the $D^0\bar{D}^0$ pair came from one $g\to c\bar{c}$ splitting. Including ISR increases the total jet yield but does not impact the ratio of $D^0\bar{D}^0$-tagged and inclusive jets on which our argument is based. The probability of having more than one $g\to c\bar{c}$ splitting in a $D^0\bar{D}^0$-tagged jet is below 1\% for $p_T^\text{jet} < 100\, \text{GeV}$.

\begin{figure}
\includegraphics[height=0.30\textwidth]{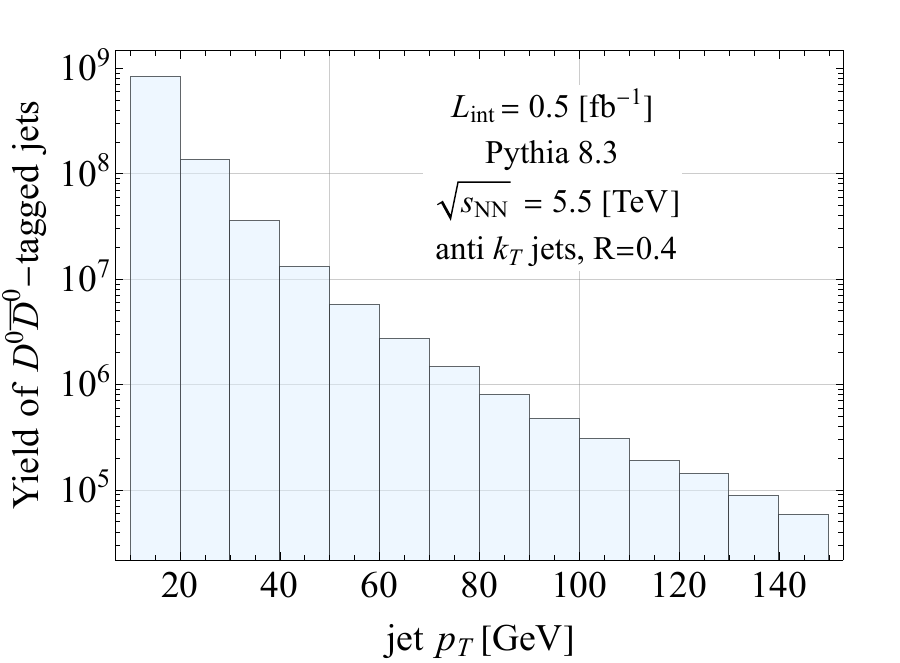}
\caption{ Expected yield of $D^0\bar{D}^0$-tagged anti-$k_t$ jets \cite{Cacciari:2008gp} with jet radius $R=0.4$ in 10 $\text{nb}^{-1}$ PbPb data, calculated  without medium-effects for an equivalent 
  ${\cal L}_\text{int}^\text{pp} =  208^2 {\cal L}_\text{int}^\text{PbPb} \approx 0.5  \, \text{fb}^{-1}$ $pp$ collisions.  }
  \label{fig3}
\end{figure}

Figure~\ref{fig3} shows the yield of jets containing a single $D^0\bar{D}^0$ pair at mid-rapidity ($\vert\eta_{\text{jet}}\vert < 1.6$) expected for the projected luminosity (${\cal L}_\text{int}^\text{PbPb} = 10\, \text{nb}^{-1}$) of the High-Luminosity Heavy-ion LHC (HL-HI-LHC)~\cite{Citron:2018lsq}. Non-prompt contributions arising from $b$-quark fragmentation can be removed experimentally and are not considered in the following. The two-body decay $D^0 \to K^- \pi^+$ can be reconstructed in ultra-relativistic heavy-ion collisions, but it has a branching ratio of $3.96 \%$~\cite{ParticleDataGroup:2020ssz} so only $1.6 \times 10^{-3}$ of all $D^0\bar{D}^0$ pairs decay via this channel. Figure~\ref{fig3} therefore indicates that ${\cal O}(10^6)$  [${\cal O}(10^3)$] counts are produced in this particular channel for $10\, \text{GeV} < p_T^{\text{jet}} < 20\, \text{GeV}$ [$80\, \text{GeV} < p_T^{\text{jet}} < 90\, \text{GeV}$], respectively. While we are not in a position to provide an experimental feasibility study for one particular detector, we conclude that there is sufficient yield to make such an effect measurable with a suitably optimized detector at the HL-HI-LHC.

  {\bf Implementation of $g \rightarrow c\bar{c}$ medium modification in parton showers.} 
To identify the $g\to c\bar{c}$ splitting in \Eq{eq1}, one requires $Q^2 \ll \hat{s}$. It is a generic feature of QCD that this scale difference $Q^2 \ll \hat{s}$ leads to a large logarithm that requires resummation. The right hand side of \Eq{eq1} is therefore written in terms of the cross section $\hat{\sigma}^{gg\to gX}$ for a gluon jet, and the $c\bar{c}$ pair can arise at any stage in the branching history of that gluon. This resummation is accounted for in a parton shower. Parton showers are simulated by evaluating branching probabilities defined in terms of parton splitting functions. 

\begin{figure*}[ht]%
\centering
\subfloat{\includegraphics[width=0.483\linewidth]{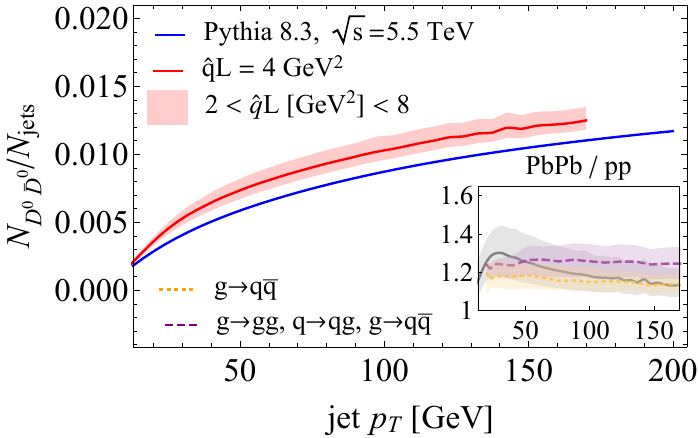}}%
\subfloat{\includegraphics[width=0.5\linewidth]{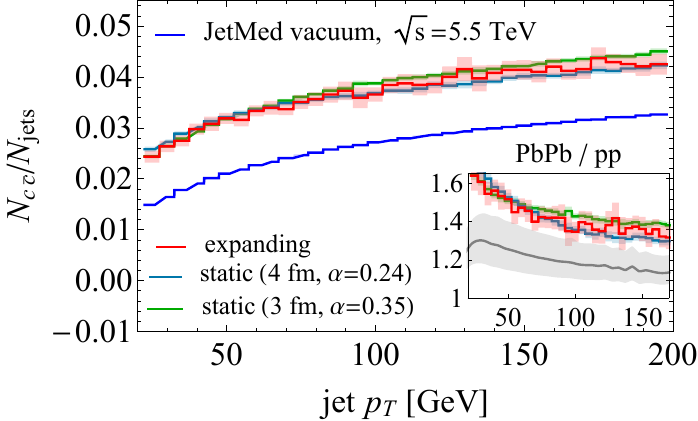}}
\caption{(a) The fraction of jets of $R=0.4$  that carry a $D^0\bar{D}^0$-tag in $\sqrt{s} = 5.5\, \text{TeV}$  mid-central ($\vert\eta_{\text{jet}}\vert < 1.6$) collisions. The \textsc{Pythia}-simulation for $pp$-collisions (blue line) is compared to medium-modifications using reweighting (Eqn.~\eqref{eq:weight}).
    (b) \texttt{JetMed} results for the fraction of jets of $R=0.4$  that carry a $c\bar{c}$-pair in $\sqrt{s} = 5.5\, \text{TeV}$  mid-central ($\vert\eta_{\text{jet}}\vert < 1.6$)
    collisions. Results from $pp$ collisions (blue line) are compared to results for in-medium propagation in static and expanding scenarios that describe the measured jet nuclear modification factors.
    Both insets show in grey the relative medium-enhancement found in this simulation.
    As seen in the inset of Fig.~\ref{fig4}(b) \texttt{JetMed} predicts a larger enhancement.
   The inset of Fig.~\ref{fig4}(a) compares to results from a simple massless dipole shower with modification of $g\to q\bar{q}$ splittings only (orange band) and with modification to  $g\to gg,q\to qg,g\to q\bar{q}$ splittings (purple band) (see Supplemental Material for details).}%
\label{fig4}%
\end{figure*}

In order to assess the medium-modifications of the $c\bar{c}$ yield in a parton shower we study three complementary implementations:

{\bf First, }we take the \textsc{Pythia} events with a $D^0 \bar{D}^0$-tagged jet described above, identify the $g\to c\bar{c}$ splitting and reweight those events according to~\cite{Attems:2022ubu} 
  \begin{equation}
  \label{eq:weight} 
    1 + 
    \frac{ \left( \frac{1}{Q^2} P_{g\to c\bar{c}} \right)^{\rm med}(E_g,\upkappa^2,z) }{\left( \frac{1}{Q^2} P_{g\to c\bar{c}} \right)^{\rm vac}(\upkappa^2,z) }\, .
  \end{equation}
  The validity of this reweighting relies on the $g \to c\bar{c}$ splitting being sufficiently rare \cite{Attems:2022ubu}. We evaluate \Eq{eq3} for an expanding medium by relating it to an equivalent static medium with adjusted $L=\tau_f-\tau_i = 4$ fm and  $ C_F \hat{\bar{q}} L = \hat{q}L = \langle q^2\rangle_\text{med}$  (see Ref.~\cite{Attems:2022ubu} for details). To account for uncertainties of this procedure and for possible model dependencies of medium-parameters~\cite{JET:2013cls,Andres:2016iys,Andres:2019eus,Caucal:2020uic,JETSCAPE:2021ehl},  we allow for a wide range $2\, \text{GeV}^2 < \langle q^2\rangle_\text{med} < 8\, \text{GeV}^2$. 
  The resulting enhancement is shown in Fig.~\ref{fig4} (a) (red) as compared to the vacuum baseline (blue).

  The advantage of this approach is that one simulates hadronic distributions and implements directly the $\bm{\upkappa}$-differential information from \Eq{eq3}. Its limitation is that the reweighting does not account for the modification of other splittings (especially $g\to gg$), which could open phase space for extra gluon splittings and lead to energy loss of the jets studied. Our assumption in this approach is hence that the extra phase space is suppressed for small medium modifications and that energy loss similarly affects inclusive and $D^0 \bar{D}^0$-tagged jets. 

{\bf Second,} we use the state-of-the-art jet quenching Monte Carlo code \texttt{JetMed}~\cite{Caucal:2020uic}. It is based on a systematic factorization of (angular-ordered) vacuum-like and medium emissions. 
The in-medium evolution is based on a Markovian iteration~\cite{Blaizot:2013vha} of the BDMPS-Z rate with flavor-dependent kernels as derived in~\cite{Arnold:2008zu,Ghiglieri:2015ala}. This implements the enhancement of the $\bm{\upkappa}$-integrated $c\bar{c}$ rate consistent with~\Eq{eq3}. The distribution of this rate in transverse phase space is chosen 
according to Brownian motion, thus modelling the main feature of the $\bm{\upkappa}$-differential distribution~\Eq{eq3}.
To include heavy quark production in \texttt{JetMed}, we enforce that $g\to c\bar{c}$ satisfies the mass threshold $\bm{\upkappa}^2\ge 4 m_c^2$ (see also Supplemental Material for details \footnote{See Supplemental Material [url] for a description of the implementation of heavy quarks in \texttt{JetMed} and additional results for the charm quark pair enhancement in a simple dipole shower, which
includes Refs.~\cite{Caucal:2018dla,Caucal:2019uvr,Caucal:2020xad,Mehtar-Tani:2018zba,Dokshitzer:1991fd,Hoche:2014rga,tutorial}.}). The advantage of \texttt{JetMed} is that it simulates the complete parton shower, including medium modifications of all parton splittings.  
We show results for a Bjorken-expanding medium (red) and for two static media with $L = 3$~fm and $L=4$~fm (green and teal, respectively) in Fig.~\ref{fig4}, with other parameters fixed to reproduce the jet nuclear modification in central PbPb collisions~\cite{Caucal:2020uic}.

{\bf Third,}  we take a simple massless vacuum dipole shower and replace all splitting functions with medium-modified versions \cite{Armesto:2009fj} with the same range of $\hat{q} L$ as in the first approach (see Supplemental Material for details). 
This implements modifications to all parton splittings with the same parameters as the first approach.  However, this does not implement hadronization or charm mass effects. Also, in contrast to \texttt{JetMed}, it is not tuned to jet spectra and their medium modifications. Its main use in the present study is to test the interplay between the modifications of different splittings.
In the inset of Fig.~\ref{fig4}(a) we show the results of this parton shower with only $g\to q\bar{q}$ enhancement (orange dotted) as well as with all splittings enhanced (purple dashed). This comparison shows that the dominant effect in the enhanced $D^0\bar{D}^0$-tagged jet yield is indeed due to the  $g\to c\bar{c}$ splitting function.

{\bf Enhancement of $g \rightarrow c\bar{c}$ in medium-modified jets.} 
All three medium-modified partons showers in Figs.~\ref{fig4} indicate an enhancement of $g\to c\bar c$ splittings compared to the vacuum baseline (blue).
 
Fig.~\ref{fig4}(a) shows the enhancement of $N_{D^0\bar{D}^0}/N_\text{jets}$ calculated via reweighting of  \textsc{Pythia}. We find a sizeable increase of $N_{D^0\bar{D}^0}/N_\text{jets}$ due to the medium-modification of $g\to c\bar{c}$. In our (simpler) third approach also $g \to g\, g $ and $q \to q g$ are medium-modified and affect $N_{D^0\bar{D}^0}/N_\text{jets}$. As seen in the inset of Fig.~\ref{fig4}(a),  the increase in the ratio $N_{D^0\bar{D}^0}/N_\text{jets}$ is dominated by $g\to q\bar{q}$. The medium-modification of $g\to gg$ and $q \to q g$ plays a smaller but non-negligible effect in this simulation.%

For the \texttt{JetMed} results in Fig.~\ref{fig4}(b), the values of the medium parameters of all scenarios are fixed to reproduce the jet nuclear modification factor. We find an enhancement of $N_{c\bar{c}}/N_\text{jets}$ that ranges from 60\% - 40\% as a function of $p_T$. $N_{D_0\bar{D}_0}$ can be obtained from $N_{c\bar{c}}$ by multiplying with the square of the $c \to D^0$ branching ratio $\approx 0.4-0.6$~\cite{ALICE:2021dhb}.

In \texttt{JetMed}, we have an independent test that the medium-induced $g\to gg$ splitting makes a negligible contribution to the enhanced $c\bar{c}$ yield. Namely, we can artificially prevent gluons produced by medium-induced radiation from splitting into $c\bar{c}$-pairs. This fixes the number of gluons available to split so that it is not enhanced by medium-induced radiation. We find that the $c\bar{c}$ enhancement in this case is not modified within line width.
This is consistent with the picture that medium-induced gluon emissions are soft and do not have sufficient virtuality to split into $c\bar{c}$-pairs.

Model-dependent differences in the three approaches are clearly visible. In particular, \texttt{JetMed} predicts an enhancement that is a factor $\approx 2$ larger than the first and third approach (see insets of Fig.~\ref{fig4}). 
This is unsurprising as the three approaches have different model parameters and rely on somewhat different assumptions. The fact that the enhancement of $N_{c\bar{c}}/N_\text{jets}$ and $N_{D^0\bar{D}^0}/N_\text{jets}$ is seen in all three cases allows us to conclude that it is a firm prediction of radiative parton energy loss.

Experimental feasibility is detector-specific and needs to be assessed within experimental collaborations. However, several general remarks can be made: First, Fig.~\ref{fig4} includes jets that contain a $D^0$ and a $\bar{D}^0$ meson of any $p_T$. Experimentally, such a measurement amounts to accessing the $ D^0\bar{D}^0$ yield down to arbitrarily soft momenta where reconstruction efficiency degrades and backgrounds are high. This, however, may not be 
 critical since the $D^0$ and $\bar{D}^0$ mesons entering Figs.~\ref{fig3} and~\ref{fig4} typically carry a rather large fraction of the jet $p_T$ (data not shown) due to the hard fragmentation of gluons into heavy quarks. In addition,
jet substructure techniques tailored to $g \to c\bar{c}$~\cite{Ilten:2017rbd} could be used in the future to access the detailed kinematics of the $g \rightarrow c\bar{c}$ splitting from the hadron level. Finally, the measurement of $N_{D^0\bar{D}^0}/N_\text{jets}$ is likely to remain statistics-limited at HL-HI-LHC due to the small branching ratio of $D_0 \to K\, \pi$. 
However, advances in tagging jets with a $c\bar{c}$ pair~\cite{Iwamoto:2017ytj} may make it possible to sample the entire $c\bar{c}$ statistics for the study of boosted
$g\to c\bar{c}$ topologies. 
 
{\bf In summary}, we have demonstrated that the standard paradigm of radiative parton energy loss implies an enhanced $g\to c\bar{c}$ splitting. We have shown that  $N_{D^0\bar{D}^0}/N_\text{jets}$ is an observable that is sensitive to this as yet unexplored, but qualitatively novel, feature of this parton energy loss paradigm. Both the magnitude of the enhancement  and the abundance of the signal make this measurement a target for the future HL-HI-LHC, where novel detector technologies will allow for the highest rates and for extreme signal purities in Run 5 and 6~\cite{ALICE:2803563}.

\textbf{Acknowledgments:} We thank Leticia Cunqueiro, Alexander Huss, Jos\'e Guilherme Milhano, Pier Francesco Monni, Andreas Morsch, Krishna Rajagopal, Alba Soto-Ontoso, Ivan Vitev, Korinna Zapp and Nima Zardoshti for useful discussions. MA acknowledges support through H2020-MSCA-IF-2019 ExHolo 898223.

\bibliography{letter}

\section{Supplemental Material}

Here, we describe in more detail the two Monte Carlo implementations of medium-modified parton showers, used in the main text. 

 {\bf Heavy quark production in \texttt{JetMed}.}
 \texttt{JetMed} implements a medium-modified parton shower
 based on a factorized picture of vacuum-like and medium-induced parton splittings. The vacuum evolution is correct at single logarithmic accuracy in the strongly ordered collinear limit~\cite{Caucal:2018dla,Caucal:2019uvr,Caucal:2020xad,Caucal:2020uic}. It is supplemented by medium-modifications as follows: 
\begin{enumerate}
\item Vacuum emissions are generated by an angular-ordered shower starting from $\theta_\text{max}=1$, with a minimal transverse momentum cut-off $k_{\perp,\text{min}}=0.25$~GeV. 
%
In the region $k_\perp^3\theta < 2\hat{q}$, medium-induced emissions are dominant  and vacuum emissions are hence vetoed. Furthermore, emission with $\theta<\theta_c=2/\sqrt{\hat{q}L^3}$ lose energy coherently with their emitter and can be treated as if they were happening outside the medium (see step~3 below). 
\item Medium-induced emissions are generated for each parton produced in the first step. The collinear splittings are ordered in an emission time $0<t<L$ and computed with multiple-soft splitting kernels~\cite{Mehtar-Tani:2018zba}.
For $g\to gg$ splitting the formation time is given by
\begin{equation}
t_f = \sqrt{\frac{2 z(1-z)xE}{\hat{q} (1-z(1-z))}}
\end{equation}
and the splitting rate is
\begin{equation}
\frac{d^2\Gamma_\text{med}}{dzdt} = \frac{\alpha_s P^\text{vac}_{a\to bc}(z)}{\sqrt{2}\pi} \frac{1}{t_f}.
\end{equation}
The coupling constant $\alpha_s$ is a fixed model parameter without scale evolution.
Partons produced both from vacuum-like and medium-induced splittings acquire an average transverse momentum
\begin{equation}
\left< k_\perp^2\right>(t,t_0) = \int_{t_0}^t dt'\hat{q}(t')
\label{inmediumkt}
\end{equation}
while traversing the medium.
The acquired transverse momentum is sampled from a gaussian distribution and added to the parton transverse momentum (which is zero for medium-induced emissions).
Emissions at angles larger than $\theta_\text{max}$ are discarded.
%
\item For each parton that leaves the medium after step 2,  vacuum-like emissions outside the medium are restarted with
$\theta_\text{max}$. That the first emission outside the medium does not obey angular ordering is a consequence of the loss of coherence in the medium.
Emissions with $k_\perp \theta> 2/L$ or $\theta>\theta_c$ are vetoed.
\end{enumerate}

The generated partons can then be passed to \texttt{FastJet} to construct infrared and collinear safe observables. By default, we consider $R=0.4$ anti-$k_\perp$ jets. 

To include charm mass effects, we have supplemented the \texttt{JetMed} evolution as follows. To logarithmic accuracy, the vacuum splitting functions $g\to q\bar{q}$ are the same for charm quarks and massless quarks, but charm quarks have to pass the mass threshold. This is implemented by vetoing $g \to c\bar{c}$ if the generated transverse momentum does not satisfy $k^2_\perp > m_c^2$.  
For medium-induced $g\to c\bar{c}$, the multiple-soft splitting kernel is the $\bm{\upkappa}$-integrated version of Eq.~3 in the main text to leading logarithmic accuracy. The pair is produced collinearly and it aquires transverse momentum in the medium as explained below \Eq{inmediumkt}. In analogy to the vacuum case, the medium-induced splitting is vetoed if the medium-induced momentum transfer is insufficient to lift the $c\bar{c}$-pair above the  mass threshold $k^2_\perp > m_c^2$. In addition, gluon radiation from a heavy quark is vetoed within the dead cone angle $m_Q/E$~\cite{Dokshitzer:1991fd}. This procedure implements mass effects to leading logarithmic accuracy in the vacuum parton shower
and treats the medium-induced $g \to c\bar{c}$ splitting on equal footing with other medium-induced splittings in \texttt{JetMed}.

 {\bf Simple dipole-shower.}
 To further establish the robustness of our main conclusions to model assumptions, we additionally provide results for a simple dipole shower supplemented with medium-modified splitting functions. We follow the public implementation of a massless $\upkappa$-ordered dipole shower with final-state radiation only~\cite{Hoche:2014rga,tutorial}. The shower is initialized with a color neutral $q\bar{q}$ or $gg$ pair of momenta
\begin{equation}
    p_1^\mu = E (1,0,1,0)\, ,\quad p_2^\mu = E (1,0,-1,0)\, ,
\end{equation}
and it is evolved in $t=\upkappa^2$ from the initial scale $E^2$ down to the cut-off scale $t_0=1\,\text{GeV}^2$. 
Gluons and $N_f=5$ massless quark flavor are included in the evolution. For the medium modification, the $q\to q g$, $g \to g g$ and (massless) $g\to q\bar{q}$ vacuum splitting functions are supplemented with additive BDMPS-Z corrections $P_{a\to b c}^\text{tot}=P_{a\to b c}^\text{vac}+P_{a\to b c}^\text{med}$. For reasons of technical simplicity, we use for all splitting functions the small-$z$ limit 
\begin{align}
  &P_{a\to b c}^\text{med}(E, \upkappa^2,z)\Big\vert_{z\ll 1} = P_{a\to b c}^\text{vac}(z) \mathfrak{I}\left(\frac{\upkappa^2}{\hat{q}_c L},\frac{zE}{\frac{1}{2}\hat{q}_c L^2}\right)\Big\vert_{z\ll 1} \, . \label{eq:modfactor}
\end{align}
Here, $\mathfrak{I}$ is the universal modification factor 
\begin{align}
&\mathfrak{I}\left(\tilde{\upkappa}^2, \tilde \omega\right)
		= \frac{\kappa^2}{2\omega^2} \, \mathfrak{Re}\, \int_0^{L} dt \int_t^{\infty} d\bar{t} \int d{\bf r}\,
    e^{ - \epsilon |t| - \epsilon |\bar{t}|}\,		\nonumber \\
		 & \times \, 
     e^{ - \frac{1}{4} \int_{\bar{t}}^\infty d\xi\, \hat{q}(\xi)\, {\bf r}^2}\, 
		  e^{-i\, \bm{\upkappa} \cdot{\bf r}}
\frac{\partial}{\partial {\bf x}}\cdot \frac{\partial}{\partial {\bf r}}
		   \, {\cal K}\left[{\bf x}=0,t;{\bf r},\bar{t}\right]\, ,
\end{align}
which depends on the rescaled variables $\tilde{\upkappa}^2 \equiv \tfrac{\upkappa^2}{\hat{q}_c L}$ and
$\tilde \omega\equiv \tfrac{\omega}{\frac{1}{2}\hat{q}_c L^2}$ with $\omega = z E$. The variable $z$ denotes the momentum fraction of the softer splittee, which is the gluon for $q\to q g$ and $g \to g g$ and the quark or anti-quark for $g\to q\bar{q}$. Correspondingly, $\hat{q}_c =\hat{q}_A = C_A\, \hat{\bar{q}}$ for  $q\to q g$, $g \to g g$ and $\hat{q}_c = C_F\, \hat{\bar{q}}$ for $g\to q\bar{q}$. 

For the $g\to gg$ and $q\to q g$ splitting functions, medium modifications of the form \Eq{eq:modfactor} have been implemented previously in the Q-\textsc{Pythia} medium-modified parton shower~\cite{Armesto:2009fj}. Our simple dipole-shower parallels this logic but includes $g\to q\bar{q}$.

\begin{figure}
\includegraphics[width=\columnwidth]{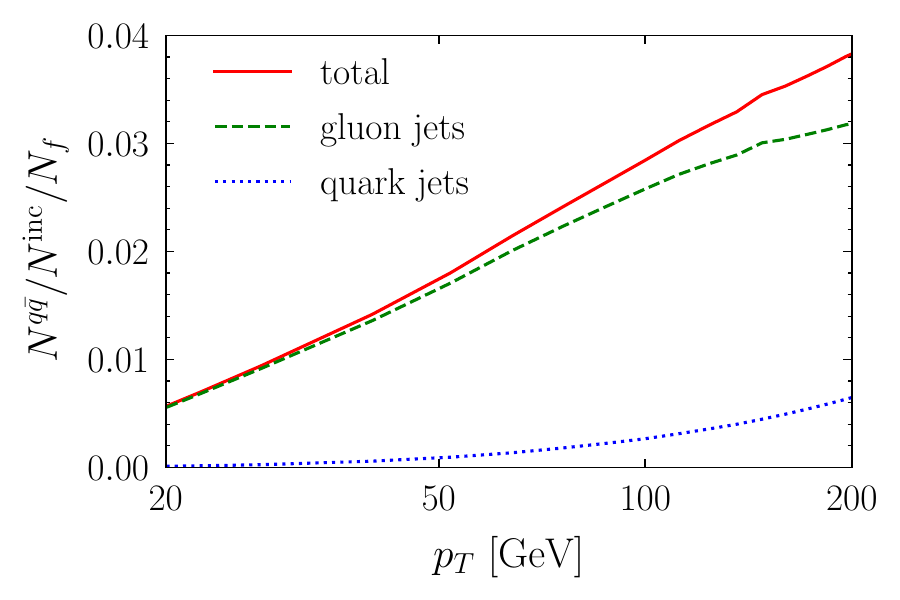}
  \caption{The yield of $q\bar{q}$-tagged $R=0.4$ jets per flavor as a fraction of all jets, calculated from the simple stand-alone dipole vacuum parton shower supplemented with \Eq{eq9}.}
\label{Afig1}
\end{figure}

{\bf Supplementary results for the simple medium-modified dipole shower.}
For each event we reconstruct anti-$k_t$ jets with $R=0.4$. We define $q\bar{q}$-tagged jets as jets containing a single $q\bar{q}$ pair that originates from the same gluon splitting. 
To simulate the fraction $N^{q\bar{q}}/N^{\rm inc}/N_f$ of the number of $q\bar{q}$-tagged over inclusive jets of specific flavor, we require additional information about the hard production rate of partons $i$ (quark or gluon) with initial energy $E$. For the relevant partonic cross-sections ${d\sigma_i}/{dE}$ we use leading order results 
averaged over rapidity window $y<1$ at $\sqrt{s}=5.02\,\text{TeV}$ from Ref.~\cite{Huss:2020whe} and we take $E=p_T$. The jet $p_T$ spectra are then given by weighting average contributions from single initial parton jets  with the respective leading-order partonic cross-section
\begin{equation}
    \frac{d\sigma_j}{dp_T} = \int dE \left( \frac{d\sigma_g}{dE} \frac{dN_{g\to j}(E)}{dp_T} +\frac{d\sigma_q}{dE}\frac{dN_{q\to j}(E)}{dp_T}\right)\, .
    \label{eq9}
\end{equation}
For the ratio $N^{q\bar{q}}/N^{\rm inc}/N_f$ of $q\bar{q}$-tagged over inclusive jets, this modeling leads to a vacuum baseline shown in \Fig{Afig1}. 

 In comparison to the vacuum baseline obtained from \texttt{JetMed} (see the blue line in Fig.~\ref{fig3}(b)), the vacuum baseline of the ratio of $q\bar{q}$-tagged over inclusive jets shown here is a factor 1.2  (1.1) larger for $p_T = 200$ (100) GeV jets. In the region of $p_T = 20 - 50$ GeV, deviations between the baselines of \texttt{JetMed} and this simple dipole shower grow larger than a factor 2. We note in this context that a dipole-shower and an angular-ordered shower agree only up to leading logarithmic accuracy. Also, the effect of charm mass is beyond leading logarithmic accuracy and it is included in \texttt{JetMed} but not in this simple dipole shower. In general, contributions beyond leading logarithmic accuracy can grow large if the phase space for parton splitting becomes small which is the case for low $p_T < 50$ GeV. Taking this into account, we conclude that the vacuum baselines of \texttt{JetMed} and of this simple dipole-shower agree within expected uncertainties.

 \begin{figure}
\centering
    \subfloat[]{
        \includegraphics[width=\columnwidth]{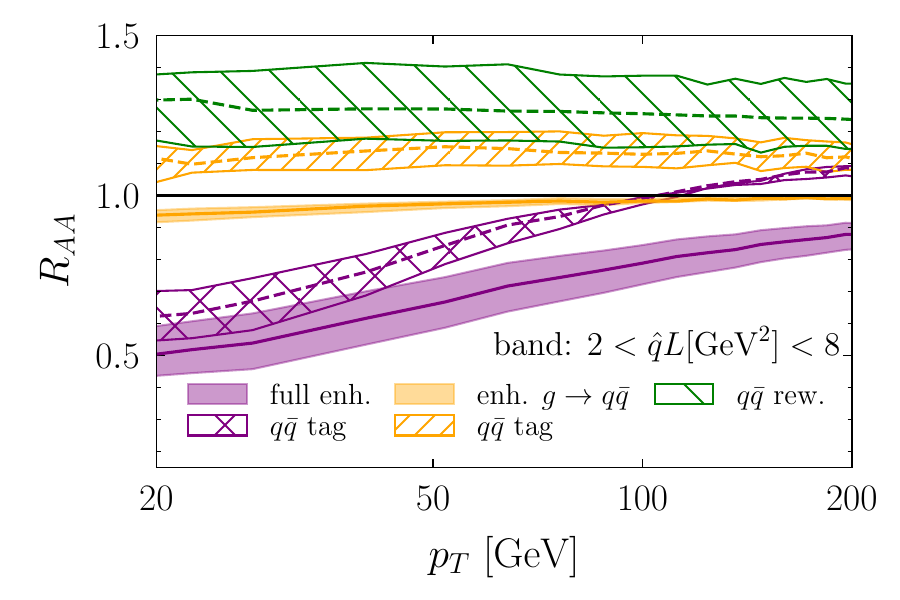}
    }

    \vspace{0.5em}

    \subfloat[]{
        \includegraphics[width=\columnwidth]{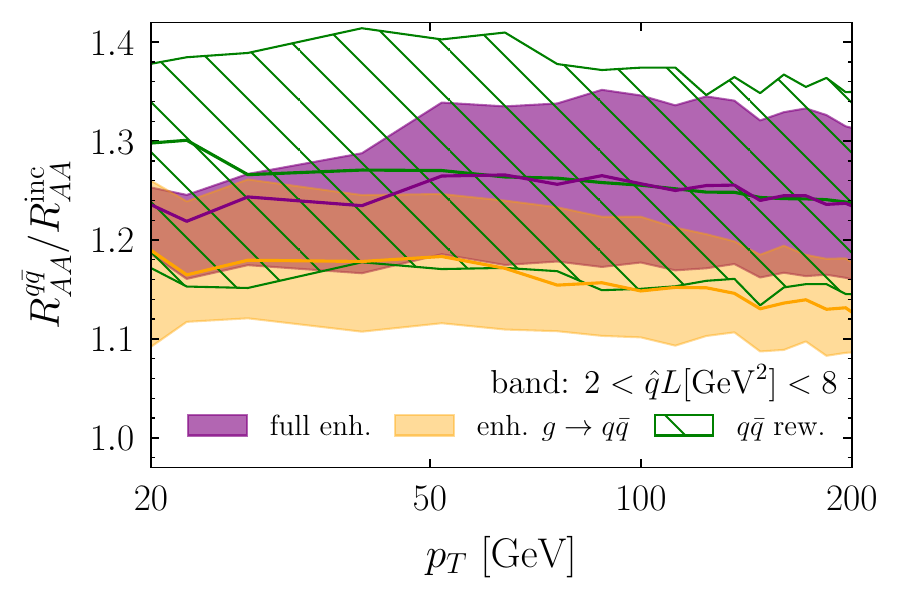}
    }
  \caption{ (a) Results of the simplified stand-alone parton shower for the nuclear modification factor of inclusive (solid band) and of $q\bar{q}$-tagged (hatched band) $R=0.4$ jets. The case of BDMPS-Z medium-modifications to all splitting functions (purple) is compared to the case of including $g\to q\bar{q}$ splitting function only (orange). Results of the reweighting procedure \eqref{eq:weight} are included (green). (b) The ratio of tagged over inclusive jet $R_{AA}$'s.
 }
\label{Afig3}
\end{figure} 

We show the nuclear modification factor $R_{AA}$ in this simulation for inclusive and for $q\bar{q}$-tagged jets in \Fig{Afig3}(a). 
If the medium modification of all splitting functions is included, then $R_{AA}^{\rm inc}$ for inclusive jets shows the significant suppression characteristic of jet quenching (purple solid). The 
$R_{AA}^{q\bar{q}}$ for  $q\bar{q}$-tagged jets shows
a significantly smaller suppression (purple hatched) since the effects leading to jet quenching in $R_{AA}^{\rm inc}$ are partially compensated by a characteristic rate enhancement due to medium-modified $g\to q\bar{q}$. If only the $g \to q\bar{q}$ splitting function is medium-modified, there is little modification of inclusive jets (solid orange) but substantial enhancement of $q\bar{q}$-tagged jets (hatched orange).
In \Fig{Afig3}(b) we show the double ratio $ R_{AA}^{q\bar{q}}/R_{AA}^{\rm inc}$, which helps to isolate the effects from medium-enhanced $q\bar{q}$ production. Irrespective of whether the parton shower includes medium modifications to all splitting functions (purple band) or only modifications to the $g\to q \bar{q}$ splitting function (orange band), a significant enhancement of the double ratio is observed. 
This signals the dominant role of medium-enhanced $g\to q \bar{q}$ splitting in this enhancement.

In both panels of \Fig{Afig3} we additionally show in green the results of the reweighting procedure applied to this parton shower. The reweighting is in reasonable agreement with the effect of enhancing only the $g \to c\bar{c}$ splitting.

\Fig{Afig3} corroborates our main conclusions in several ways. First, within the bands of $\hat{q}L$ variation, the enhancement of $R_{AA}^{q\bar{q}} / R_{AA}^{\rm inc}$ obtained from the medium-modified dipole parton shower is comparable to the enhancement obtained from the reweighting procedure, thus justifying the use of the latter in the main paper. Second, 
inclusion of medium-modified $g\to gg$ and $q\to q g$ splitting has a numerically small effect on the ratio $R_{AA}^{q\bar{q}}/R_{AA}^{\rm inc}$ (cf.~\Fig{Afig3}(b)) although it affects the nuclear modification factors $R_{AA}^{\rm inc}$ and $R_{AA}^{q\bar{q}}$ significantly (cf.~\Fig{Afig3}(a)). 
This supports our conclusion that the $g\to q\bar{q}$-induced enhancement of $q\bar{q}$-tagged jets can be identified as an enhancement of $N_{D^0\bar{D}^0}/N_{\rm jets}$ even in the presence of energy loss. It is also consistent with the conclusion reached in a different way in our \texttt{JetMed} study (see main text) that the enhanced $c\bar{c}$-radiation is not due to medium-enhanced $g\to gg $ followed by vacuum splitting of $g\to c \bar{c}$, but that it is due to medium-enhanced $g\to c\bar{c}$ splitting.

\end{document}